\documentclass[aps,pre,reprint,twocolumn,showpacs,amsmath,amssymb]{revtex4-1}

\usepackage{graphicx}
\usepackage{bm}
\usepackage{color}

\begin{document}

\title{Splash control of drop impacts with geometric targets}

\author{Gabriel Juarez}
\thanks{Author to whom correspondence should be addressed.}\email{gjuarez@seas.upenn.edu}

\author{Thomai Gastopoulos}

\author{Yibin Zhang}

\author{Michael L. Siegel}

\author{Paulo E. Arratia}

\affiliation{Department of Mechanical Engineering and Applied Mechanics,
University of Pennsylvania, Philadelphia, PA 19104, USA}

\date{\today}

\begin{abstract}

Drop impacts on solid and liquid surfaces exhibit complex dynamics due to the 
competition of inertial, viscous, and capillary forces. After impact, a liquid lamella 
develops and expands radially, and under certain conditions, the outer rim breaks up 
into an irregular arrangement of filaments and secondary droplets. We show experimentally 
that the lamella expansion and subsequent break up of the outer rim can be controlled by 
length scales that are of comparable dimension to the impacting drop diameter. Under 
identical impact parameters, ie. fluid properties and impact velocity, we observe 
unique splashing dynamics by varying the target cross-sectional geometry. These behaviors 
include: (i) geometrically-shaped lamellae and (ii) a transition in splashing stability, 
from regular to irregular splashing. We propose that regular splashes are controlled by
the azimuthal perturbations imposed by the target cross-sectional geometry and that 
irregular splashes are governed by the fastest-growing unstable Plateau-Rayleigh mode.

\end{abstract}


\maketitle

\section{Introduction}

The impact of liquid drops is a rich phenomenon that continues to draw copious research 
attention since drop impacts are ubiquitous to many processes in both nature and industry 
\cite{Worthington, Edgerton, Peregrine, deGennes, Prosperetti, Bergeron, Bergeron01}. Ink-jet 
printing, pesticide deposition, and fuel combustion are just a few examples where the 
effective application of a fluid onto a surface relies on the impact and subsequent splash 
of drops. Despite the fascination with splashing patterns \cite{Marmanis, Thoroddsen}, the
dominant mechanism that leads to the rim break up, filament formation, and secondary droplets 
remains controversial \cite{Allen, Yarin95, Zhang}.

Recently, a better understanding of how to influence splashing, ie. either enhance or 
suppress the occurrence of a splash, has been obtained. Drop impacts under different 
carefully chosen experimental conditions, such as on compliant surfaces \cite{Pepper}, 
on moving surfaces \cite{Bird}, on wetted patterned surfaces \cite{Lee}, in environments 
of varying pressure and gas composition \cite{Xu}, and with non-Newtonian liquids \cite{Bergeron}
has provided techniques that can precisely control splashing. The dominant mechanism,
however, still remains unclear. One reason for the ambiguity is that for all of the 
above cases, the length scale of the target surface is much larger than the impacting 
drop diameter. Under such conditions, the impact process is defined by the competition 
of inertial, viscous, and capillary forces \cite{Rein, Yarin}. Unfortunately, it is 
difficult to distinguish the role played by each force, and as a result, it has been 
challenging to formulate reliable theoretical and numerical methods.

\begin{figure}[b]
\centering
	\includegraphics[width=\linewidth]{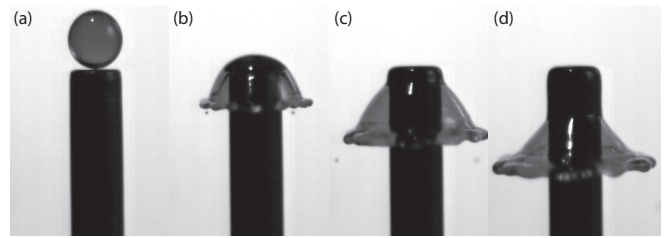}
\caption{Side view of drop impact on a cylindrical post recorded at 40 000 fps. (a) A drop
of diameter 2.85 mm  with impact velocity of 1.56 m s$^{-1}$ makes contact with the target.
The drop deforms and (b) and spreads radially to form (c) a liquid lamella sheet. (d) As the 
sheet expands, undulations along the rim emerge followed by the formation of filaments and 
secondary smaller droplets. The time interval between frames is equal to the characteristic
impact time, $\tau^{\ast} \approx 1.8$ ms. \texttt{See Supplemental Material at [URL inserted 
by publisher] for movie} \cite{SuppMat}.}
	\label{fig:sideview}
\end{figure}

In this manuscript we provide insight into the instability governing the break up of 
liquid lamella sheets that develop after drop impact. Liquid drops of diameter $D_0$ 
fall onto a target post of equal diameter with impacting speed $U_0$. A finite amplitude
azimuthal perturbation is produced by varying the target cross-sectional geometry, 
which includes a cylinder and regular polygon shapes. Figure~\ref{fig:sideview} 
shows the side view of an example drop impact with a cylindrical post with a time 
interval between frames in terms of the characteristic impact time, $\tau^* = D_0/U_0$. 
Despite the advantage of this simple setup, only a limited number of investigations 
have focused on drop impacts with obstacles of similar length scales as a window to 
understanding the complexities of drop splashing \cite{Rozhkov02, Rozhkov03, Rozhkov04, 
Josserand, Bakshi, Subramani, Villermaux}.

\begin{figure*}
\centering
	\includegraphics[width=\linewidth]{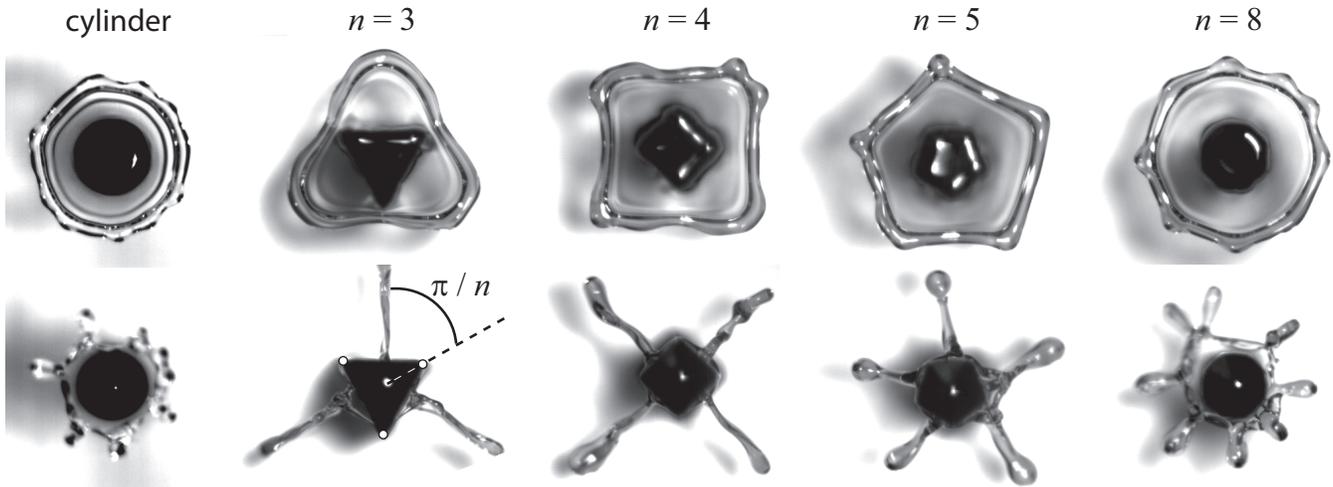}
\caption{Top view of drop impact on geometric target posts. (Top row) Geometrically-shaped 
lamella $2\tau^*$ after impact with a cylindrical, triangular ($n=3$), square ($n=4$), pentagon 
($n=5$), and octagon ($n=8$) post. For $n < 8$, the resulting lamella shapes are identical to 
the target geometry but are rotated by $\pi/n$ due to the azimuthal variation of viscous dissipation. 
(Bottom row) Filament formation $4\tau^*$ after impact shows that the splashing dynamics depend 
on the target cross-sectional geometry. The lamella rims for $n=3$, 4, and 5 break up in a controlled 
manner and form the exact number of filaments as the number of target vertices. The lamella rims 
for the cylinder post and $n \geq 8$ targets, however, break up in a similar fashion independent 
of target shape. \texttt{See Supplemental Material at [URL inserted by publisher] for movies}
\cite{SuppMat}.}
	\label{fig:topview}
\end{figure*}

\section{Experimental methods}

Droplets are created as liquid is injected into a capillary tube using a low-noise syringe 
pump. The liquid slowly drips out of the tube to form reproducible pendant drops with an 
average diameter $D_0$ of $2.85$ mm. The liquid is composed of de-ionized water and glycerol.
Food coloring is added to the solution for image enhancement purposes. The liquid has a 
viscosity of $10$ cP and a surface tension with ambient air of $35.3\times 10^{-3}$ N m$^{-1}$. 
Drops fall from a height of $15$ cm before striking the target, hitting the surface with a 
measured impact velocity $U_0$ of $1.56$ m s$^{-1}$. All experiments are performed at ambient 
pressure (101 kPa). The dynamics are described by two dimensionless parameters; the Reynolds 
number (Re), defined as $\rho D_0 U_0 / \mu$, and the Weber number (We), defined as $\rho D_0 
U_0^2 / \gamma$. Here, $\rho$ is the fluid density, $U_0$ is the impact velocity, $D_0$ is 
the drop diameter, $\mu$ is the dynamic viscosity, and $\gamma$ is the surface tension. 
For the given set of experimental parameters, this results in a Re of 550, ie. inertial
forces dominate viscous forces, and a We of 250, ie. inertial forces dominate surface forces. 
The capillary number, defined as $\mu U_0/\gamma$, is 0.45 meaning that surface forces dominate 
over viscous forces. Top and side view images are recorded using high-speed photography 
ranging from 30 000 to 40 000 fps.

The target posts are machined out of polyoxymethylene with no surface treatments. The target
cross-sectional geometry is varied and includes a cylinder and regular polygon shapes that 
range from a triangle ($n=3$) to a decagon ($n=10$), where $n$ is the number of vertices. 
The diameter of the cylindrical post is $2.85$ mm, equal to the impacting drop diameter, and 
the impacting cross-sectional surface area is kept constant for all shapes (cylinder and polygons)
at $6.38$ mm$^2$. This geometric constraint allows the polygonal circumradius, the radius 
of a circle that passes through all of the polygon vertices, to be expressed in terms of 
the initial drop diameter as a function of the number of vertices given by
\begin{equation}
\mathcal{R}(n) = D_0 \sqrt{\frac{\pi}{2n \sin(2\pi/n) }} \ .
\end{equation}
More importantly, the relevant azimuthal length scale, which is the edge length between vertices, 
is given by
\begin{equation}
s(n) = 2 \mathcal{R}(n)\sin(\pi/n) \ .
	\label{eqn:sidelen}
\end{equation}
From equation~\eqref{eqn:sidelen}, we note that the edge length is largest for $n=3$ and 
decreases as the number of vertices increase. This effectively decreases the amplitude of 
the azimuthal perturbation.

\section{Results}

\subsection{Effect of target cross-section on drop impacts}

Figure~\ref{fig:topview} shows snapshots from the top view of a drop impacting target posts of 
different cross-sectional geometries. Under similar impacting conditions, ie. constant Reynolds 
and Weber numbers, we observe that the spreading and retraction of the liquid lamella is 
significantly affected by the target cross-sectional geometry. For example, both regular 
($3\leq n <8$) and irregular (cylinder and $n\geq 8$) splashing is observed for impacts on 
polygonal posts. We refer to regular splashing as whenever the number of filaments is equal 
to the number of target vertices and their location is rotated azimuthally by an angle of 
$\pi/n$ with respect to the target orientation. Irregular splashing occurs when the number 
of filaments that form, and their location, are independent of the target geometry, or number 
of vertices.

For the cylindrical case, the drop deforms and spreads radially upon impact (Figs.~\ref{fig:sideview} 
and~\ref{fig:topview}). A thick rim forms at the edge of the lamella sheet due to the accumulation 
of ejected fluid. As the rim decelerates due to surface tension, it becomes susceptible to 
infinitesimal perturbations that lead to the break up of the lamella sheet into filaments and 
secondary droplets. As the cross-sectional geometry of the post is changed, the dynamics of the 
resulting lamella are significantly altered. Figure~\ref{fig:topview} (top row) shows example 
snapshots of geometric lamella for $n=3$, 4, and 5 at a time $2\tau^*$ after impact, where $\tau^*$ 
is the characteristic impact time. Strikingly, the resulting splash resembles the shape of the 
polygonal target with an azimuthal rotation of approximately $\pi/n$ with respect to the target 
orientation, where $n$ is the number of vertices. For example, a drop that impacts a triangular 
post results in a triangular-like splash that is shifted by $\pi/3$ with respect to the post 
(Fig.~\ref{fig:topview}, $n=3$). For $n \geq 8$, the splashing dynamics are similar to the 
cylindrical post case.

\subsection{Dynamics of geometrically-shaped lamella}

\begin{figure}
\centering
	\includegraphics[width=8cm]{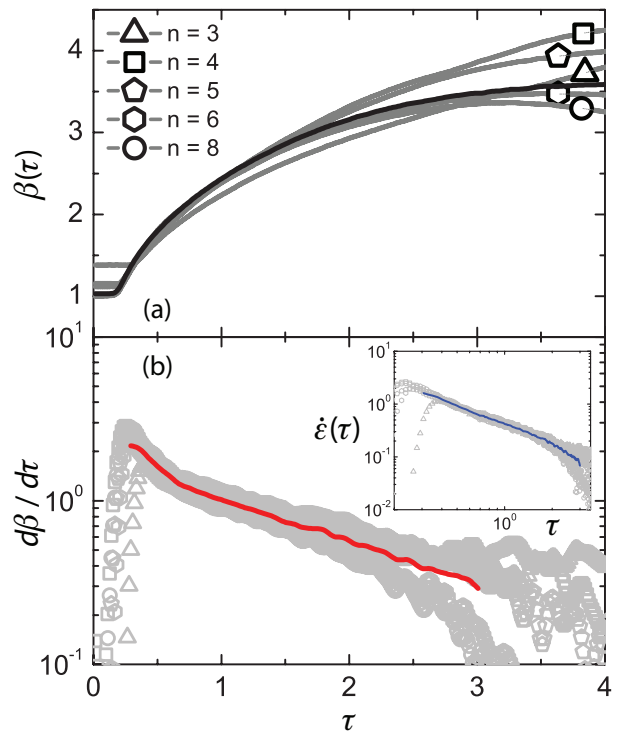}
\caption{(Color online) (a) The lamella splash diameter normalized by the initial drop 
diameter plotted as a function of normalized time $\tau$. The average maximum splash 
diameter for all targets is $3.74$ which agrees well with both scaling laws of $\beta_m \sim$ 
Re$^{1/5}$ and $\beta_m \sim$ We$^{1/4}$. (b) The velocity of the expanding splash 
diameter exhibits two exponentially decaying regimes. At early times ($0.3 < \tau < 0.7$), 
a fast decay is due to the inertia dominated deformation of the drop as it contacts 
the target. At later times ($0.7 < \tau < 3$), a second slower decay is due to 
viscous dissipation and surface forces impeding lamella expansion. (Inset) The average 
strain rate on the expanding lamella sheet, $\dot{\varepsilon} = \dot{\beta}/\beta$, shows 
two exponential regimes.}
	\label{fig:characterization}
\end{figure}

The dynamics of lamella sheets are characterized by measuring the normalized splash 
diameter $\beta$, which is the ratio of the instantaneous splash diameter $D(\tau)$ 
and the initial drop diameter $D_0$, as a function of normalized time $\tau = t/ \tau^*$
(Fig.~\ref{fig:characterization}a). Here, $\tau = 0$ is taken to be the instant 
that the drop makes contact with the surface of the target. The first 
few instants, as the lamella spreads along the target surface from the point 
of impact, are not able to be resolved and represent the initial flat part of $\beta(\tau)$.
Each plot of $\beta(\tau)$ represents an average of at least five impact events.
The maximum normalized splash diameter $\beta_m$ for all target cross-sections is 
$3.74 \pm 0.33$. The average value of the maximum normalized splash diameter agrees 
reasonably well with the scaling laws of $\beta_m \sim$ Re$^{1/5}$ and with $\beta_m \sim$ 
We$^{1/4}$ \cite{Clanet}. This means that inertia, viscous, and surface forces play 
important roles in the splashing dynamics despite the minimal interaction between 
the drop and the target surface. This is in accordance with an impact number $P\equiv$ 
We/Re$^{4/5}$ close to unity \cite{Clanet}. Values of $P<1$ describe impacts for 
inviscid fluids and $P>1$ describe impacts of viscous fluids. For this study, the 
impact number is $P \approx 1.6$ and therefore follows closely with both scaling laws.

The liquid lamella expansion rate, computed from the splash diameter $d\beta/d\tau$, shows
two exponentially decaying regimes (Fig.~\ref{fig:characterization}b). At early times 
($0.3 < \tau < 0.7$), the rim expansion follows a fast decay due to the inertia dominated 
deformation of the drop as it comes into contact with the target. The initial downward 
momentum is transferred horizontally, producing radial expansion parallel to the surface 
of the target. At later times ($0.7 < \tau < 3$), the rim expansion is described by a 
second slower decay that than first regime. Viscous dissipation is present due to shear 
flow at the target surface as well as surface forces due to the increase in surface area, 
both working to impede the lamella expansion. For $\tau < 0.3$, the rapid increase in 
the expansion rate is due an artifact as the initial transient of spreading along the 
target surface is not captured until the lamellae expand beyond the target circumradius. 
The corresponding lamella strain rate $\dot{\varepsilon}$, computed here as the ratio 
of the expansion rate $d\beta / d\tau$ and the normalized splash diameter $\beta(\tau)$, 
shows two exponential regimes in accordance with biaxial extensional flow. This would 
suggest that the splashing dynamics could be very different for non-Newtonian fluids 
where the extensional viscosity can vary by orders of magnitude under strong 
extensional flows \cite{Bergeron, Rozhkov03}.

As noted earlier, the resulting splash resembles the target polygonal shape but with 
an azimuthal rotation with respect to the target orientation (eg. Fig.~\ref{fig:topview}, $n=3$).
The rotation of the lamella by $\pi/n$ relative to the target can result from two possible
mechanisms: (i) the rapid decrease in kinetic energy as the drop deforms after impact and
(ii) the azimuthal dependence of viscous dissipation in the boundary layer that is formed
in the vicinity of the target surface. Let us consider a geometric cross-sectional target
that is described by the smallest and the largest radial distance from the origin, the
apogee $r$ and the circumradius $\mathcal{R}$, respectively. For a liquid drop that expands
radially in contact with the surface from the origin, the time it takes for the liquid 
lamella to reach the apogee is less than the time it takes to reach the circumradius. The 
fluid at the apogee experiences less of a decrease in kinetic energy and less viscous 
dissipation than the fluid at the circumradius. Hence, the fluid velocity is larger at 
the apogee than at the circumradius resulting in a geometrical lamella that are shifted 
by $\pi/n$ with respect to the target vertices, for $n<8$.

\subsection{Rim instability: regular and irregular splashing}

\begin{figure*}
\centering
	\includegraphics[width=\linewidth]{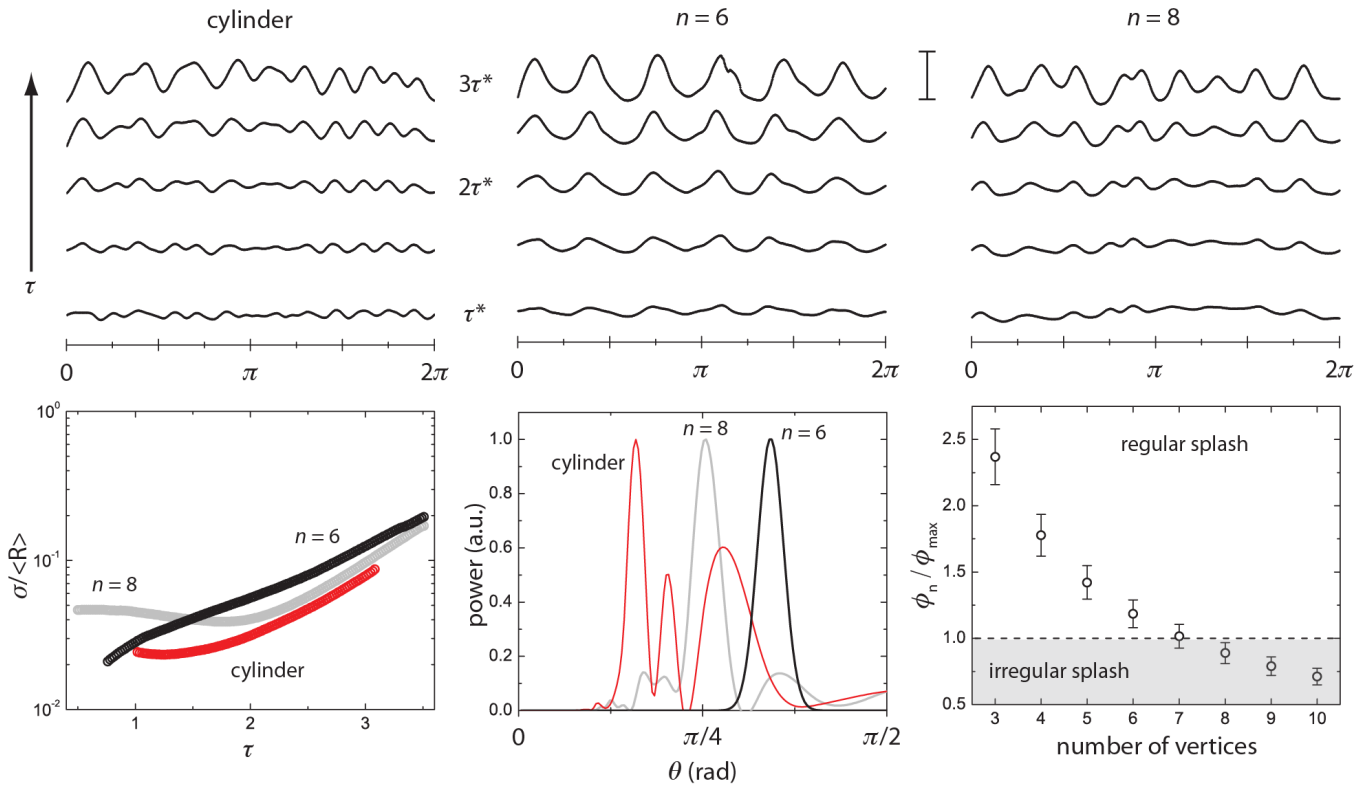}
\caption{(Color online) (Top row) Evolution of the radial profile for expanding lamella 
sheets for the time interval $\tau^* < \tau < 3\tau^*$ after drop impact on a cylinder, 
hexagon ($n=6$), and octagon ($n=8$) target. Six equidistant peaks are evident over the 
entire interval for the hexagon case. The peaks for the cylinder and octagon case are not 
evenly spaced and the number decreases due to merging. Scale bar represents 1.75 mm. 
(Bottom row) \textit{Left:} The deviations in the undulation amplitude $\sigma$ normalized 
by the average lamella radius $\langle R \rangle$ increases exponentially with time. 
\textit{Middle:} The periodograms of the radial profiles of the lamella sheets for the 
three cases at $2\tau^*$ after impact. There is a single narrow peak at $\pi/3$ for the 
$n=6$ case and a broad distribution of values with multiple peaks for the cylinder and 
$n=8$ cases. \textit{Right:} A comparison of the target perturbation amplitude $\phi_n$ 
and the most unstable azimuthal mode $\phi_{\textrm{max}}$ of a toroid jet determined 
by the PR instability. Regular splashing occurs when $\phi_n / \phi_{\textrm{max}} > 1$
for targets with $3 \leq n < 8$ and irregular splashing occurs when $\phi_n / \phi_{\textrm{max}} 
< 1$ for targets with $n \geq 8$.}
	\label{fig:radialFFT}
\end{figure*}

Once the maximum splash diameter is reached, the liquid lamella retracts inward. 
Finger formation and secondary droplets result as the outer rim breaks up in 
order to minimize the increase in surface energy. We observe that for polygonal targets, 
the ability to create geometric lamellae that undergo controlled break up into $n$ 
filaments depends on the target cross-sectional geometry and holds for targets with 
$n < 8$ only (Fig.~\ref{fig:topview}, bottom row). Specifically, there is a transition 
in the splashing stability from regular ($3 \leq n < 8$) to irregular ($n \geq 8$) break 
up of the liquid lamellae. We propose that there is a competition between the finite 
amplitude perturbation imposed from the target cross-sectional geometry and the most 
unstable mode determined by the dominant instability, which in this case is similar 
to the Plateau-Rayleigh (PR) instability \cite{Eggers97, Rozhkov02, Zhang}. Other 
possible mechanisms that have been proposed include the Richtmyer-Meshkov \cite{Gueyffier} 
and the Rayleigh-Taylor \cite{Thoroddsen, Allen, Krechetnikov, Villermaux} instabilities.

In order to gain insight into the mechanism responsible for the break up and retraction 
of the lamella, an analysis of the corrugations around the expanding rim was performed. 
The top row of figure~\ref{fig:radialFFT} shows the evolution of the azimuthal profile of 
lamella sheets after drop impact on a cylinder (left), hexagon (center), and octagon (right) 
target over the time interval of $\tau^*$ to $3\tau^*$. At early times ($\tau = \tau^*$), 
the amplitude of rim undulations is similar for all three cases. At later times however 
($\tau^* < \tau \leq 3\tau^*$), it is evident that the radial profile for the hexagon case 
is different than the profiles for the cylinder and octagon cases. Typical behavior of 
lamella sheets for impacts on targets with $3 \leq n<8$ is that there are $n$ equidistant 
peaks apparent over the entire splash process, similar to the six equidistant peaks for 
the hexagon case. For other targets ($n \geq 8$) the peaks are unevenly distributed and 
the number decreases as filaments merge during the sheet expansion, similar to the profiles 
for the cylinder and octagon cases (Fig.~\ref{fig:radialFFT}, top row). 

For all cases ($3 \leq n \leq 10$ and cylinder), the amplitude of rim undulations 
increase with time. The fluctuations of the corrugations, which are quantified by the ratio 
of the standard deviation $\sigma$ about the average lamella sheet radius $\langle R \rangle$, 
grow exponentially with time (Fig.~\ref{fig:radialFFT}, bottom left). The rates of growth,
evident by the slope of the straight portion of the curves for $\tau > 2$, are similar for 
all cases independent of the target cross-sectional geometry. This is not surprising because 
the mechanism behind every lamellae break up, whether it undergoes regular or irregular 
splashing, is driven by surface tension. The exponential growth rate, however, is indicative 
of a PR instability. The dispersion relation associated with the most unstable mode of the 
PR instability is given by $\omega_{\textrm{PR}} = 0.34 \sqrt{ \gamma / \rho a^3 }$,
where $\gamma$ is the surface tension, $\rho$ is the fluid density, and $a$ is the radius 
of the fluid jet \cite{Lhuissier}. For the current experimental parameters, the time scale
of the PR instability would be approximately $2.7$ ms. The average characteristic time scale 
of growth in fluctuations of expanding lamellae, extracted by fitting an exponential function 
to the curve for $\tau > 2$, is measured to be $2.1 \pm 0.4$ ms, in good agreement with the
PR time scale.

For further comparison, we compute the periodograms of the radial profiles for the three 
cases (cylinder, $n=6$, and $n=8$) at $2\tau^*$ (Fig.~\ref{fig:radialFFT}, bottom middle). 
The periodogram of the radial profile for the hexagon case is a single narrow peak centered 
about $\pi/3$. It is typical for the periodograms to contain a single peak centered about 
$\pi / n$ for targets with $3 \leq n < 8$ vertices. The periodograms for other targets 
(cylinder and $n \geq 8$), however, are broad and contain multiple peaks, represented in 
the periodograms for the cylinder and octagon cases. This supports the idea that the 
perturbation imposed from the target cross-sectional geometry for $3 \leq n < 8$ 
overwhelms the most unstable mode and is therefore a determining factor in the evolution 
of the lamella. The cylinder and octagon cases, however, contain a distribution of values 
as the imposed target perturbation is small compared to the most unstable mode, making 
the rims unstable and susceptible to infinitesimal perturbations.

It seems reasonable to conclude that a regular splash will occur when the azimuthal 
perturbation imposed by the target cross-sectional geometry is larger than the most 
unstable mode of the expanding toroidal jet, experimentally equivalent to the thick 
outer rim. The thin liquid lamella sheet that connects the outer rim to the target post 
is neglected since we believe that it does not contribute to the rim instability. This 
simplification is supported by our observations that, for moderate Re, there are no ripples 
in the lamella sheet (Fig.~\ref{fig:topview}) as seen for high Re impacts of $\mathcal{O}(10^4)$ 
\cite{Rozhkov02}. Furthermore, the lamellae are seen to break from the outer most 
points of the rim rather than from within the sheet connecting the target to the rim.

Utilizing the observations that the fluctuations in the rim corrugations increase 
exponentially with a characteristic time similar to that associated with the PR dispersion
relation $\omega_{\textrm{PR}}$, we approximate the most unstable mode of a toroidal jet as 
determined by the PR instability and compare it to the azimuthal perturbation imposed due to 
the target geometry. The rim volume can be expressed as a fraction of the initial drop volume 
$V_r = \varepsilon V_0$, with $0 < \varepsilon < 1$. Denoting $a$ as the minor radius of the 
toroid, $R_m$ as the major radius of the toroid, and $R_0$ as the initial radius of the 
impacting drop, the rim volume is given by
\begin{equation}
2 \pi^2 a^2 R_m = \frac{4}{3} \pi \varepsilon R_0^3 \ .
\end{equation}
At maximum expansion, the torus minor radius can be written in terms of the maximum splash
radius $R_m$ and the normalized splash radius $\beta_m$ and is given by
\begin{equation}
a = R_m \sqrt{ \frac{2 \varepsilon}{3 \pi \beta_m^3} } \ .
	\label{eqn:minorrad}
\end{equation}
Using the scaling relations for $\beta_m$ \cite{Clanet} and an average measured value for 
$\varepsilon$ of $0.65$ \cite{Rozhkov02, Rozhkov04}, equation~\eqref{eqn:minorrad} predicts
the minor radius $a$ to be $0.27$ mm. This value agrees well with observations of the rim 
thickness for geometric lamella, measured to be $a = 0.3$ mm.

Analogous to the most unstable wavelength of a cylindrical jet \cite{Rayleigh}, the most 
unstable azimuthal mode for a toroid jet determined by the PR instability \cite{Pairam, McGraw} 
is given by 
\begin{equation}
\phi_{\textrm{max}} = \frac{\lambda_{\textrm{max}}}{R_m} = \frac{9.02 \ a}{R_m} = 9.02 \ \sqrt{
\frac{2 \varepsilon}{3 \pi \beta_m^3} } \ .
	\label{eqn:phimax}
\end{equation}
The amplitude of the azimuthal perturbation imposed for regular polygon targets is taken 
to be 
\begin{equation}
\phi_n = \pi / n \ .
	\label{eqn:phitarget}
\end{equation}
The ratio of the target perturbation $\phi_n$ and the most unstable azimuthal mode of a 
toroid jet $\phi_{\textrm{max}}$ is plotted as a function of target vertices $n$ 
(Figure~\ref{fig:radialFFT}, bottom right). Interestingly, we see that for targets with
$3 \leq n < 8$ vertices, the azimuthal perturbation imposed by the target geometry is 
larger than the most unstable azimuthal PR mode, or that $\phi_n / \phi_{\textrm{max}} > 1$. 
These conditions will produce a regular splash, in agreement with observations (Fig.~\ref{fig:topview}).
For targets with $n \geq 8$, however, the azimuthal perturbation is smaller than the 
most unstable PR mode, $\phi_n / \phi_{\textrm{max}} < 1$, suggesting that the lamella 
rim is susceptible to infinitesimal perturbations and will produce an irregular splash,
independent of target geometry. We note that $R_m$ does not explicitly appear in 
equation~\eqref{eqn:phimax} as we assume that the maximum radius for the toroid jet is 
similar to circumradius of a geometrically-shaped lamella, a reasonable assumption from 
Fig.~\ref{fig:characterization}(a). Finally, to show that these results are independent 
of scaling arguments, both $\beta_m \sim$ We$^{1/4}$ and $\beta_m \sim$ Re$^{1/5}$ are 
used in place of the normalized splash radius in equation~\eqref{eqn:phimax}, and the 
upper and lower bounds are shown with error bars.

\section{Conclusions}

We have shown that the expansion and subsequent break up of the outer rim of 
liquid lamellae can be controlled by length scales on the order of the impacting 
drop diameter. Under identical impact conditions of constant Reynolds and Weber 
numbers, we observe unique splashing dynamics by simply varying the target cross-sectional
geometry to include a cylinder and regular polygon shapes. For polygon targets with 
$3 \leq n < 8$ vertices, the expanding lamellae resemble the geometric cross-section 
of the target, but are rotated by an angle of $\pi/n$ with respect to the target 
orientation. We find that the break up of the outer rim and liquid lamellae are well 
controlled and reproducible. The number of filaments that form during splashing is 
equal to the number of vertices $n$ of the target. For other targets (cylinder and 
$n \geq 8$), the expansion and break up of the outer rim and liquid lamellae are 
independent of the target geometry.

We find that there are two distinct splashing regimes depending on the number of 
target vertices, regular splashing ($3 \leq n < 8$) and irregular splashing (cylinder 
and $n \geq 8$). We propose that the transition in splashing stability is a result 
of the competition between the amplitude of the azimuthal perturbation imposed by 
the target cross-sectional geometry and the most unstable azimuthal mode, determined 
by the Plateau-Rayleigh instability, of the expanding outer rim. For $3 \leq n < 8$ 
polygon targets, regular splashing occurs since the imposed target perturbation is 
large enough to overwhelm the most unstable mode and effectively control the dynamics 
of the splash. For the cylinder and $n \geq 8$ targets, irregular splashing occurs 
since the imposed target perturbation is smaller than the most unstable mode and the 
resulting splash dynamics are independent of the target geometry. The rim dynamics are 
instead governed by the most unstable azimuthal Plateau-Rayleigh mode.

In summary, we show that drop splashing can be potentially controlled by the target
geometric features. The experiments presented here provide a new method that systematically 
probes the effect of azimuthal perturbations to expanding lamellae after drop impact.
While our experimental observations indicate that the splashing phenomenon is dominated
by the Plateau-Rayleigh instability, questions still remain. One important parameter to 
investigate further is the dependence of the ratio of the maximum splash radius to the 
minor radius of the outer rim, expressed in equation~\eqref{eqn:minorrad}, on varying 
impact conditions, ie. changing both the Re and the We. This would provide a better 
understanding on the limiting case for irregular splashing of liquid lamellae.

\section*{Acknowledgements}

We thank D. Hu, D. Lohse, N. C. Keim, V. Garbin, X. N. Shen, and M. Garcia for helpful 
discussions. We also thank P. Rocket for fabricating the regular $n-$sided polygon 
target posts. This work was partially supported by the National Science Foundation
through the award CBET-0932449.

\bibliography{drop_impact_geometry}

\end{document}